\documentclass{article}
\usepackage{spconf,amsmath,graphicx,hyperref}
\usepackage[nolist]{acronym}
\usepackage{adjustbox}
\usepackage{amssymb}  
\usepackage{booktabs}
\usepackage{multirow}
\usepackage{makecell}
\usepackage{xcolor}
\usepackage{subfig}

\newacro{TSE}[TSE]{Target Speaker Extraction}
\newacro{ASR}[ASR]{Automatic Speech Recognition}
\newacro{OCR}[OCR]{Optical Character Recognition}
\newacro{pTSET}[pTSE-T]{Presentation Target Speaker Extraction with unaligned Text cues}
\newacro{TPE}[TPE]{Text Prompt Extractor Network}
\newacro{TSR}[TSR]{Text-Speech Recognition Network}
\newacro{SNR}[SNR]{Signal-to-Noise Ratio}
\newacro{DPRNN}[DPRNN]{Dual-Path Recurrent Neural Network}
\newacro{ASD}[ASD]{Active Speaker Detection}
\newacro{SER}[SER]{Speech Emotion Recognition}
\newacro{BSS}[BSS]{Blind Speech Separation}
\newacro{SDR}[SDR]{Signal-to-Distortion Ratio}
\newacro{SDRi}[SDRi]{Signal-to-Distortion Ratio improvement}
\newacro{SI-SDR}[SI-SDR]{Scale-Invariant Signal-to-Distortion Ratio}
\newacro{SI-SDRi}[SI-SDRi]{Scale-Invariant Signal-to-Distortion Ratio improvement}
\newacro{LLM}[LLM]{Large Language Model}
\newacro{SNR}[SNR]{Signal-to-Noise Ratio}
\title{pTSE-T: Presentation Target Speaker Extraction using Unaligned Text Cues}
\name{Ziyang Jiang$^{1}$, Jiahe Lei$^{1}$, Xueyan Chen$^{1}$, Yifan Zhang$^{1}$, Zexu Pan$^{2}$,
Wei Xue$^{3}$, Xinyuan Qian$^{1, *}$
\thanks{
This research is supported by the National Natural Science Foundation of China
(Grant No. 62306029), the Beijing Natural Science Foundation (Grant No. L233032),
and the Young Elite Scientists Sponsorship Program of the Beijing High Innovation Plan.
}
\thanks{* Corresponding author.}
}
\address{$^{1}$ University of Science and Technology Beijing (USTB), China \\
         $^{2}$ Tongyi Lab, Alibaba Group, Singapore \\
         $^{3}$ Hong Kong University of Science and Technology (HKUST), China}
\begin{document}

\address{$^{1}$ University of Science and Technology Beijing (USTB), China \\
         $^{2}$ Tongyi Lab, Alibaba Group, Singapore \\
         $^{3}$ Hong Kong University of Science and Technology (HKUST), China}
\ninept
\maketitle

\begin{abstract}
\ac{TSE} aims to extract the clean speech of the target speaker in an audio mixture, eliminating irrelevant background noise and speech. While prior work has explored various auxiliary cues including pre-recorded speech, visual information, and spatial information, the acquisition and selection of such strong cues are infeasible in many practical scenarios. 
Differently, in this paper, we condition the \ac{TSE} algorithm on semantic cues extracted from limited and unaligned text contents, such as condensed points from a presentation slide. This method is particularly useful in scenarios like meetings, poster sessions, or lecture presentations, where acquiring other cues in real time may be challenging.
To this end, we design two different networks. Specifically, our proposed \ac{TPE} fuses audio features with content-based semantic cues to facilitate time-frequency mask generation to filter out extraneous noise. The experimental results show the efficacy in accurately extracting the target speaker's speech by utilizing semantic cues derived from limited and unaligned text, resulting in SI-SDRi of 12.16 dB, SDRi of 12.66 dB, PESQi of 0.830 and STOIi of 0.150. The dataset and source code will be publicly available at \url{https://slideTSE.github.io/}
\end{abstract}
\vspace{-0.3cm}

\section{Introduction}
\label{sec:intro}

\vspace{-0.3cm}

The `cocktail party problem'~\cite{cherry1953some}, where multiple speakers talk concurrently amidst noise, poses a significant challenge for computational speech processing tasks like \ac{ASR}~\cite{yue2019end} and \ac{SER}~\cite{pan2020multi}. A prominent approach to address this is \ac{TSE}, which aims to extract a specific speaker's speech from mixture by leveraging auxiliary cues, mimicking human selective auditory attention. Existing \ac{TSE} methods can be broadly categorized by their reliance on different types of cues. The most common paradigm uses a pre-recorded audio sample, including
the SpEx family~\cite{xu2020spex, spex_plus2020, ge2021multi} and X-sepformer~\cite{liu2023x} adopting this method. However, this approach's reliance on clean speech samples raises both practicality and privacy concerns~\cite{zmolikova2023neural}. Other methods leverage auxiliary hardware to capture spatial or visual cues. For instance, spatial cues from microphone arrays~\cite{gu2019neural} and visual cues from cameras, such as lip movements~\cite{pan2021muse, pan2022usev,10448398} and face movements~\cite{9858007,10814693,chung2020facefilter}, are employed, but their efficacy is often limited by the need for specialized hardware, and they struggle with non-stationary speakers or challenging conditions like occlusions and poor lighting.
To overcome these limitations, text has emerged as a more flexible cue. While some works have used text to separate general sounds~\cite{liu22w_interspeech, kilgour2022text, ma2024clapsep}, they typically do not address the challenge of disentangling multiple simultaneous speakers. Closer to our task, LLM-TSE~\cite{hao2023typing} utilizes text for speaker extraction, but it crucially requires an exact, pre-aligned transcript of the target speech, a prerequisite that is rarely met in real-world scenarios. Therefore, a critical gap remains, as existing \ac{TSE} methods are often constrained by specialized hardware, pre-recorded speech samples, or precisely aligned transcripts. Our work addresses this gap by proposing a universal approach that relies solely on a text prompt, eliminating the need for such restrictive conditions.
To overcome these limitations, we explore a ubiquitous yet underutilized scenario: presentations where speakers use visual aids like slides. The text on these slides, while not temporally aligned with the speech, provides strong semantic cues about the spoken content. In this work, we introduce a novel task: \textbf{presentation-based Target Speaker Extraction with Text (pTSE-T)}, as illustrated in Fig.~\ref{fig:introduction}. This task uses unaligned semantic text from presentation slides to extract the corresponding presenter's speech from a mixed audio signal. Our key contributions can be summarized as follows:
\begin{figure}[!t]
 \centering
 \includegraphics[width=\linewidth]{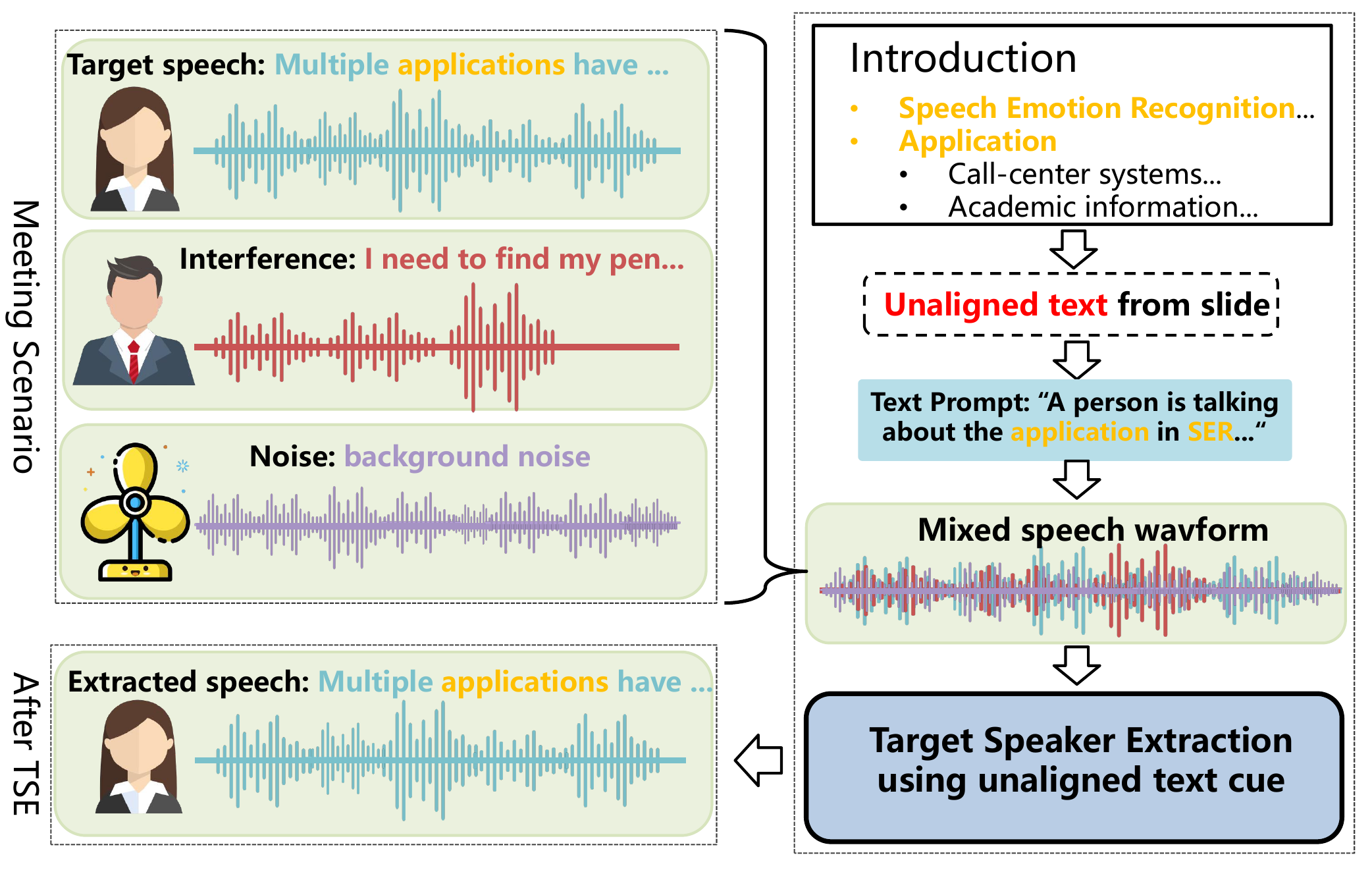}
 \caption{Illustration of our proposed pTSE-T task which extracts the presenter's speech from the audio mixture using unaligned text from the visual presentation slide.}
 \label{fig:introduction}
 \vspace{-0.6cm}
\end{figure}

\begin{figure*}[!tb]
  \centering
  \includegraphics[width=\linewidth]{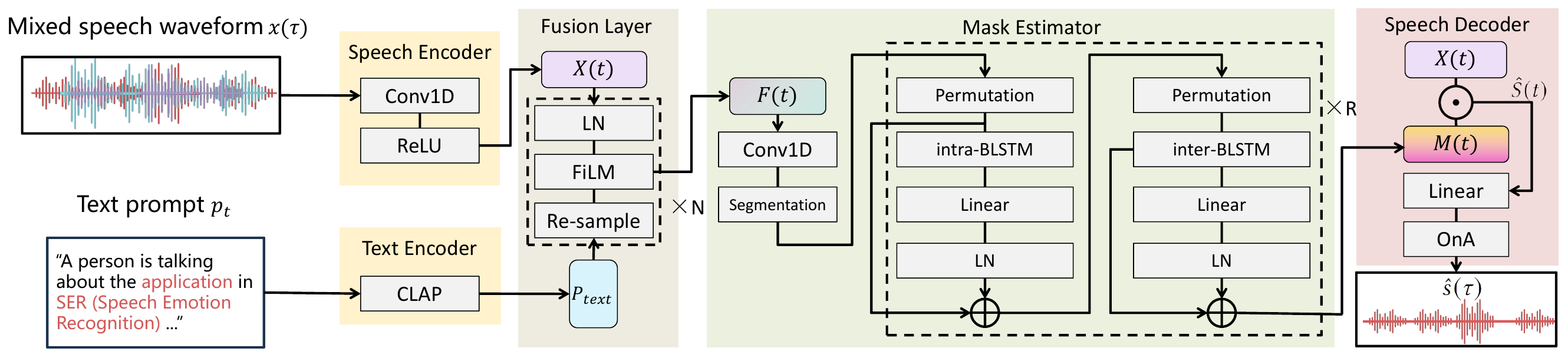}
  \caption{The structure of our proposed \ac{TPE} network. The input consists of mixed speech waveform $x(\tau)$ and pre-processed text prompts $p_t$, which are separately processed by a speech encoder and a text encoder to obtain $X(t)$ and $P_{text}$. Subsequently, within the fusion layer, we apply $\text{FiLM}(\cdot)$ for hierarchical fusion of features from both modalities to obtain $F(t)$. The fused features are then processed by a mask estimator to generate an estimated mask $M(t)$, which is element-wise multiplied with $X(t)$ and passed through the speech decoder to obtain the predicted speech waveform $\hat{s}(\tau)$ of the target speaker.} \vspace{-0.4cm}
  \label{fig:TPE}
\end{figure*}

\begin{enumerate}
    \item We are the first proposal of the \ac{pTSET} task, as illustrated in Fig.~\ref{fig:introduction}, which extracts the presenter's clean speech from the audio mixture given the unaligned semantic text cues obtained from the presentation slide.
    
    \item To facilitate this new \ac{pTSET} task, we design and annotate an innovative data corpus, namely MMSpeech, which includes temporally synchronized speech and presentation slides. In addition, we also provide the \ac{ASR} transcripts and the text descriptions extracted from slides using an open-source \ac{OCR} toolkit.
    \item We propose a novel network, i.e., \ac{TPE}. Specifically, \ac{TPE} fuses the text prompts with audio features for \ac{TSE}, achieving correct extraction rates of 96.46\% and 95.26\% on the MMSpeech-2mix and -3mix datasets, respectively.
\end{enumerate}

\vspace{-0.4cm}
\section{Proposed Methods}

We denote $s(\tau)$ as the target speech, and $\alpha_i$ as the scalar variable used to adjust the \ac{SNR} for interfering speech $v_i(\tau)$, where $i\in \{1,2,...,I\}$, $I$ denotes the number of interfering speakers and $\tau \in \{1,2,...,T\}$ denotes the time index. The mixed speech is defined as 
$x(\tau) = s(\tau) + \sum_{i=1}^I \alpha_i v_i(\tau)$,
where $s(\tau)$ is the target speech, $\alpha_i$ adjusts the \ac{SNR}, and $v_i(\tau)$ is the $i$-th interference.

Our model's design is directly motivated by the unique challenges of the proposed pTSE-T task. The primary challenge is the lack of temporal alignment between the slide text and the speech signal. This requires the model to: \textbf{1)} effectively fuse high-level semantic information with a complex audio mixture, and \textbf{2)} capture long-range temporal dependencies to associate this semantic content with the correct speaker over extended periods. Our architecture, detailed below, addresses these challenges through a specialized fusion module and a long-context mask estimator.

As illustrated in Fig.~\ref{fig:TPE}, the end-to-end network consists of five components. First, to bridge the modality gap, a \textbf{speech encoder} (1-D Conv) transforms the waveform $x(\tau)$ into a latent feature embedding $X(t) \in \mathbb{R}^{D\times T}$, while a pre-trained and frozen CLAP model~\cite{wu2023large} serves as the \textbf{text encoder}. We chose the CLAP model because its understanding of text-audio relationships provides a powerful semantic prior, crucial for interpreting the unaligned text prompt $p_t$ and mapping it to a meaningful embedding $P_{\text{text}}$.

The core challenge lies in conditioning the audio stream on these unaligned semantic cues. A simple feature concatenation would be insufficient. To solve this, our \textbf{fusion layer} uses Feature-wise Linear Modulation (FiLM)~\cite{perez2018film}. This allows text embedding $P_{\text{text}}$ to dynamically generate affine transformation parameters (scale and shift) that modulate speech features $X(t)$. This process effectively guides the network to focus on acoustic patterns relevant to the slide's content. This conditioning is applied hierarchically across $N$ blocks to refine the fused representation $F(t)$ at multiple feature scales.

Furthermore, because the text cue is semantic and not tied to a specific moment, the \textbf{mask estimator} must analyze long-term audio context to identify the correct speaker. To this end, we adopt the \ac{DPRNN} architecture~\cite{luo2020dual}. Its ability to model long-term dependencies is critical for this task. By segmenting the feature sequence into overlapping chunks and applying iterative intra- and inter-chunk processing, the DPRNN can effectively integrate local phonetic details with the global topic-level context needed to estimate the final time-frequency mask $M(t)$.

Finally, the \textbf{speech decoder} applies the estimated mask to the speech embedding ($\hat{S}(t) = M(t) \odot X(t)$) and reconstructs the target waveform $\hat{s}(\tau)$ via a transposed convolution. The entire network is trained end-to-end using the negative \ac{SI-SDR} loss. The loss is defined as:
\begin{equation}
    \mathcal{L}_{\text{TPE}} = -20 \log_{10} \left( \frac{\left\| \frac{\langle\hat{s}, s\rangle s}{\|s\|^2} \right\|}{\left\| \hat{s} - \frac{\langle\hat{s}, s\rangle s}{\|s\|^2} \right\|} \right)
\end{equation}

\vspace{-0.3cm}

\section{Experiments}
\vspace{-0.2cm}
\begin{figure}[!tb]
  \centering
  \includegraphics[width=.9\linewidth]{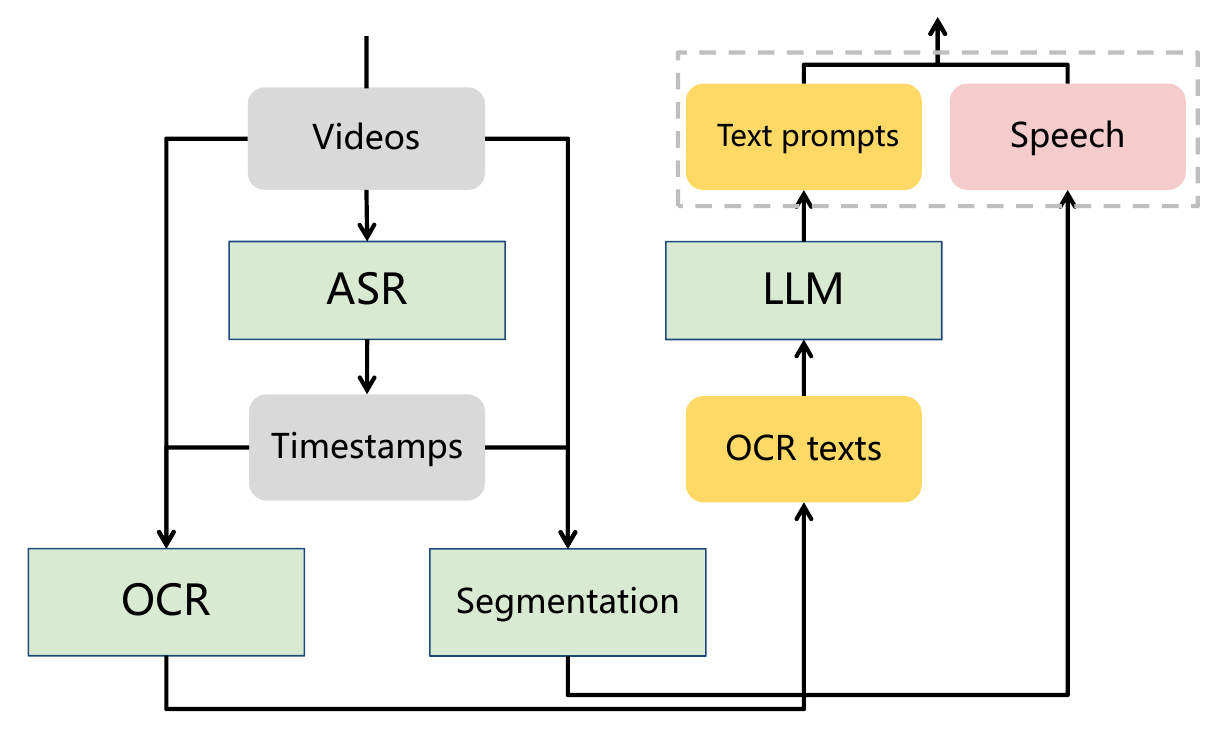}

  \caption{The data processing pipeline for constructing our MMSpeech dataset. The dashed rectangle is the input of our \ac{TSE} network.}
  \vspace{-0.5cm}
  
  \label{fig:datapipeline}
  
\end{figure}
\subsection{MMSpeech Dataset}

\renewcommand{\arraystretch}{0.9}
\begin{table*}[!t]
    \centering
    \captionsetup{skip=2pt}
    \caption{The experimental results of speech separation and our proposed \ac{TSE} networks (`$-$' denotes information not applicable).}
    \begin{adjustbox}{max width=0.9\textwidth}
    \setlength{\tabcolsep}{3.5pt} 
    \begin{tabular}{l l rrrrr rrrrr}
        \toprule\toprule
        & & \multicolumn{5}{c}{\textbf{MMSpeech-2mix (-noisy)}} & \multicolumn{5}{c}{\textbf{MMSpeech-3mix (-noisy)}} \\
        \cmidrule(lr){3-7} \cmidrule(lr){8-12}
        Task & Network & SI-SDRi$\uparrow$ & SDRi$\uparrow$ & PESQi$\uparrow$ & STOIi$\uparrow$ & Acc(\%)$\uparrow$ & SI-SDRi$\uparrow$ & SDRi$\uparrow$ & PESQi$\uparrow$ & STOIi$\uparrow$ & Acc(\%)$\uparrow$ \\
        \midrule
        Speech & DPRNN (PIT) & 14.96 & 15.18 & 0.930 & 0.176 & $-$ & 11.51 & 11.79 & 0.472 & 0.245 & $-$ \\
        separation & DPRNN (PIT, Noisy) & 13.92 & 14.18 & 0.900 & 0.180 & $-$ & 11.02 & 11.39 & 0.409 & 0.230 & $-$ \\
        \midrule
         & DPRNN (random) & -3.92 & -2.52 & -0.008 & -0.203 & 50.73 & -13.15 & -9.09 & -0.340 & -0.203 & 17.93 \\
         & LASS (w/o fine.) & -7.14 & -5.06 & -0.663 & -0.203 & 49.95 & -4.54 & -2.98 & -0.289 & -0.121 & 34.26 \\
        Speaker & CLAPSep (w/ fine., w/ neg) & -2.62 & -1.67 & -0.177 & -0.088 & 49.90 & -2.43 & -1.57 & -0.191 & -0.063 & 32.07 \\
        extraction & AudioSep (w/ fine.) & 8.59 & 9.06 & 0.577 & 0.111 & 90.20 & 5.21 & 5.54 & 0.332 & 0.098 & 70.79 \\
        & LLM-TSE (w/ fine., w/o ref) & 10.97 & 11.65 & 0.540 & 0.123 & 93.60 & 10.30 & 10.06 & 0.563 & 0.147 & 90.26 \\
        \cmidrule{2-12}
        & \bfseries Our TPE & \bfseries 12.16 & \bfseries 12.66 & \bfseries 0.830 & \bfseries 0.150 & \bfseries 96.46 & \bfseries 11.94 & \bfseries 12.41 & \bfseries 0.791 & \bfseries 0.194 & \bfseries 95.26 \\
        \bottomrule\bottomrule
    \end{tabular}
    \end{adjustbox}
    \label{table:result}
\end{table*}
\vspace{-0.3em}

\renewcommand{\arraystretch}{0.9}
\begin{table*}[!tb]
    \centering
    \captionsetup{skip=2pt}
    \caption{SI-SDR results of TPE on different noise datasets and SNR levels (trained with WHAM!).}
    \begin{adjustbox}{max width=0.9\textwidth}
    \begin{tabular}{c c c c c c c c c c c c}
        \toprule\toprule
        \multirow{2}*{Noise dataset} & \multicolumn{11}{c}{SI-SDR (dB) $\uparrow$ across different SNR (dB) levels} \\
        \cmidrule{2-12}
          & SNR=5 & SNR=4 & SNR=3 & SNR=2 & SNR=1 & SNR=0 & SNR=-1 & SNR=-2 & SNR=-3 & SNR=-4 & SNR=-5 \\
        \midrule
        \textcolor{gray}{WHAM!~\cite{Wichern2019WHAM}} & \textcolor{gray}{11.61} & \textcolor{gray}{11.25} & \textcolor{gray}{10.84} & \textcolor{gray}{10.39} & \textcolor{gray}{9.86} & \textcolor{gray}{9.24} & \textcolor{gray}{8.48} & \textcolor{gray}{7.66} & \textcolor{gray}{6.75} & \textcolor{gray}{5.81} & \textcolor{gray}{4.77} \\
         ESC-50~\cite{piczak2015dataset} & 10.74 & 10.29 & 9.77 & 9.22 & 8.61 & 7.97 & 7.25 & 6.45 & 5.61 & 4.67 & 3.72 \\
         MS-SNSD~\cite{reddy2019scalable} & 11.40 & 10.97 & 10.50 & 9.97 & 9.37 & 8.70 & 7.98 & 7.20 & 6.33 & 5.40 & 4.39 \\
         MUSAN-noise~\cite{snyder2015musan} & 10.72 & 10.25 & 9.73 & 9.17 & 8.56 & 7.93 & 7.26 & 6.50 & 5.69 & 4.81 & 3.83 \\
         MUSAN-music~\cite{snyder2015musan} & 11.05 & 10.59 & 10.08 & 9.49 & 8.82 & 8.06 & 7.24 & 6.31 & 5.29 & 4.19 & 3.03 \\
         Nonspeech~\cite{rashid2023nonspeech7k} & 10.74 & 10.27 & 9.74 & 9.15 & 8.51 & 7.83 & 7.11 & 6.32 & 5.49 & 4.58 & 3.62 \\
         QUT-NOISE~\cite{dean2010qut} & 10.88 & 10.43 & 9.91 & 9.35 & 8.70 & 7.96 & 7.14 & 6.26 & 5.22 & 4.07 & 2.79 \\
         UrbanSound8k~\cite{salamon2014dataset} & 10.99 & 10.56 & 10.06 & 9.49 & 8.87 & 8.15 & 7.35 & 6.47 & 5.53 & 4.49 & 3.35 \\
        
        \bottomrule\bottomrule
    \end{tabular}
    \end{adjustbox}
    \label{tab:SISDR_diff_noise}
\end{table*}

We introduce the \textbf{MMSpeech} dataset. It is constructed from 12.5 hours of academic presentation videos, with audio resampled to 16KHz. The dataset is collected from INTERSPEECH 2020 recordings~\footnote{INTERSPEECH 2020: http://www.interspeech2020.org}. The multimodal information is re-organized using the following procedure shown in Fig. \ref{fig:datapipeline}. The dataset provides synchronized speech clips, their \ac{ASR} transcripts, and the corresponding slide text extracted via \ac{OCR}.
First, we employ the \textbf{Whisper-large-v3} model\footnote{https://huggingface.co/openai/whisper-large-v3} to generate precise sentence-level timestamps and transcripts, which guide the segmentation of audio clips. For each clip, we extract the corresponding video frame at its starting timestamp and use the powerful \textbf{Paddle-OCR} engine\footnote{https://github.com/PaddlePaddle/PaddleOCR} to recognize the slide text. To filter out irrelevant visual elements, text bounding boxes smaller than 1\% of the frame resolution are discarded. Finally, the raw OCR output is refined into a clean, natural language prompt using the \textbf{Mistral-7B-Instruct-v0.2} model\footnote{https://huggingface.co/mistralai/Mistral-7B-Instruct-v0.2}. 
To further validate the quality, we randomly sampled 100 clips for manual inspection. For ASR, the transcripts were compared against human references, yielding an average Word Error Rate (WER) of 2.8\%. For OCR, we defined a prompt as \emph{accurate} if it preserved the key semantic information of the corresponding slide (e.g., main title, bullet points, or critical terminology). Based on this criterion, 97\% of the processed OCR prompts were judged accurate. These results demonstrate that the textual annotations in MMSpeech are of high fidelity and reliable for downstream modeling.
From the processed clips, we created the MMSpeech-2mix and -3mix datasets by mixing two or three random speech waveforms, with the target-to-interferer level difference uniformly sampled in $[-5, 5]$ dB. The dataset is partitioned into training, validation, and test sets, containing 40k (66.7h), 5k (8.3h), and 1k (1.7h) samples, respectively. It comprises 59 unique speakers, with no speaker overlap between the sets (40 for training, 10 for validation, and 9 for testing). Additionally, to evaluate robustness, we created noisy versions by adding WHAM! noise~\cite{Wichern2019WHAM} at an \ac{SNR} between -3 dB and 3 dB.

\vspace{-0.3cm}

\subsection{Implementation Details}
\vspace{-0.1cm}

The encoder of \ac{TPE} is a Conv1D with a kernel size of 16, a stride of 8, and 256 output channels. For the DPRNN block, we set the number of BLSTM layers $R=6$, window size $K=80$, and fusion layer parameter $N=2$. The hidden channels are set to 128.
We compare our model with several baselines. For \textbf{DPRNN}~\cite{luo2020dual}, we report both the separation upper bound using PIT association and the extraction performance using random association. For other baselines, including \textbf{LASS}~\cite{liu22w_interspeech}, \textbf{AudioSep}~\cite{liu22w_interspeech}, \textbf{CLAPSep}~\cite{ma2024clapsep}, and \textbf{LLM-TSE}~\cite{hao2023typing}, we fine-tuned their official implementations on our dataset. Specific setup details are as follows: for CLAPSep, we used the target text as the positive query and interfering text as the negative query, denoted as ``(w/ neg)'' in Table~\ref{table:result}. For LLM-TSE, we only used the text prompt for guidance without the reference speaker audio, denoted as ``(w/o ref)''. We finetune LLM-TSE using the same configuration of the LLaMA-2 7B model~\footnote{LLaMA-2 7B model: \url{https://huggingface.co/meta-llama/Llama-2-7b}} and the LoRA Adapter, but replace the original text prompt with our own.
Our proposed TPE network was trained using the Adam optimizer with an initial learning rate (LR) of $5\times 10^{-4}$. We employed a learning rate scheduler that halves the LR if the validation loss does not improve for 2 consecutive epochs, with early stopping after 10 epochs (min LR of $1\times 10^{-8}$). For all baseline models, we followed the training configurations (e.g., optimizer, learning rate) described in their respective original papers to ensure a fair comparison.

\begin{figure*}[t]
    \centering
    \subfloat[\ac{SI-SDR}=-8.31 dB]{\includegraphics[width=0.16\textwidth]{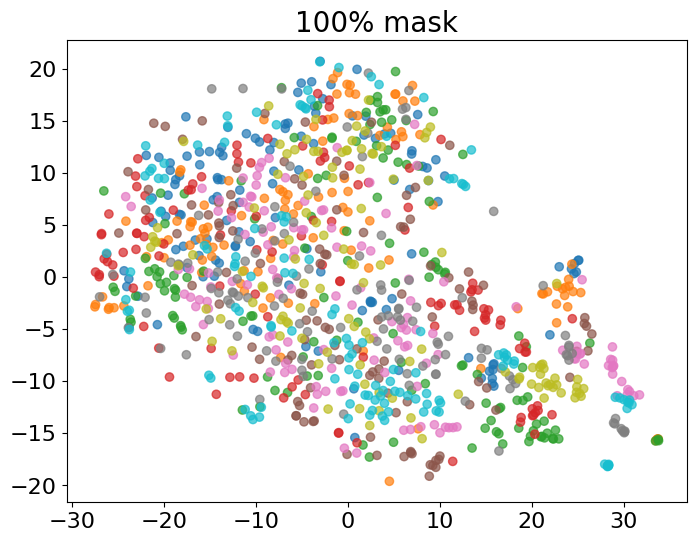}}%
    \subfloat[\ac{SI-SDR}=-0.77 dB]{\includegraphics[width=0.16\textwidth]{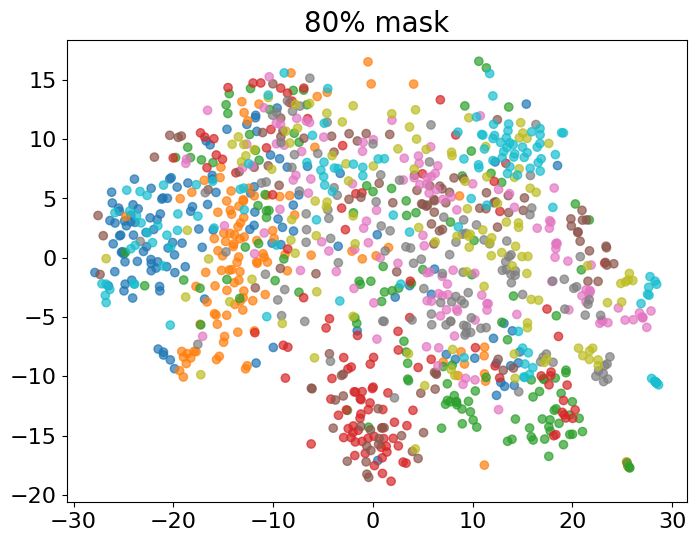}}
    \subfloat[\ac{SI-SDR}=4.42 dB]{\includegraphics[width=0.16\textwidth]{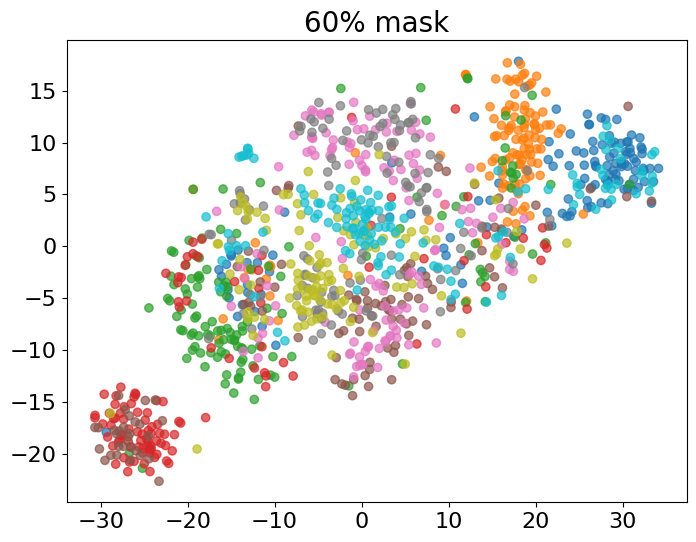}}
    \subfloat[\ac{SI-SDR}=9.11 dB]{\includegraphics[width=0.16\textwidth]{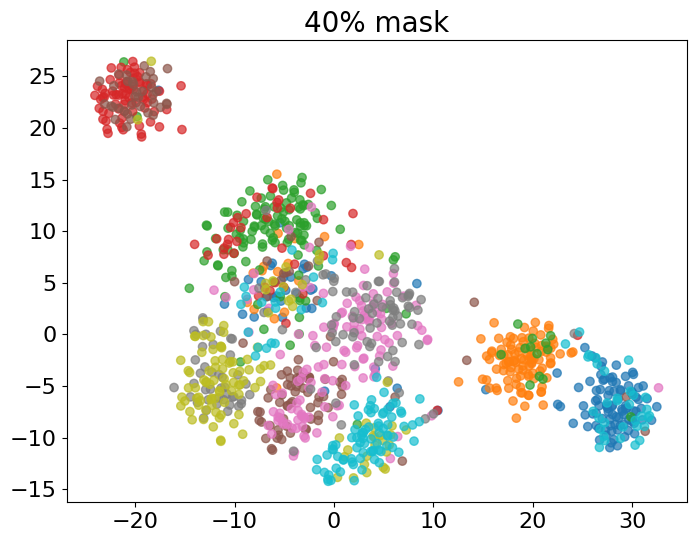}}
    \subfloat[\ac{SI-SDR}=11.05 dB]{\includegraphics[width=0.16\textwidth]{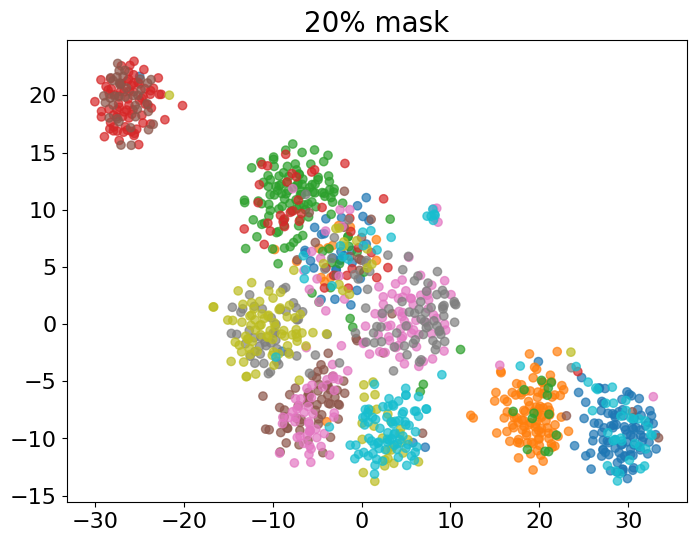}}%
    \subfloat[\ac{SI-SDR}=12.82 dB]{\includegraphics[width=0.16\textwidth]{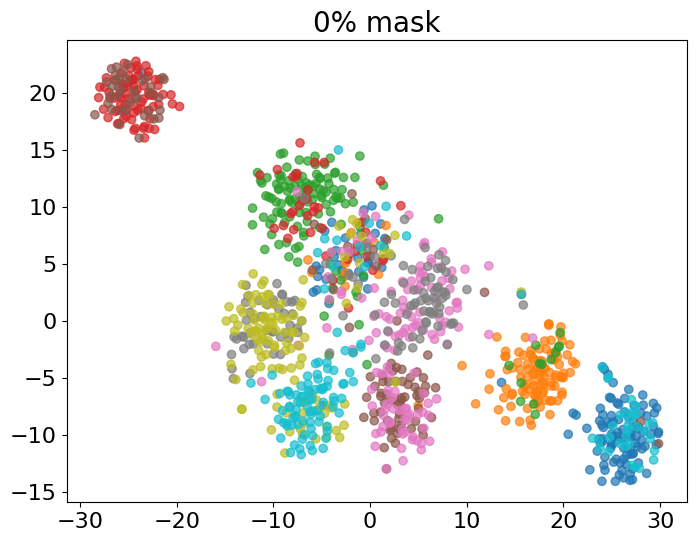}}
    \caption{The t-SNE visualization of the output of the final layer of the DPRNN block. Points with the same color indices represent the same speaker ID.}
    \vspace{-0.5cm}
    
    \label{fig:text_mask}
\end{figure*}

\vspace{-0.3cm}

\subsection{Results}

\vspace{-0.1cm}

We perform all experiments using the MMSpeech-2 (3) mix and MMSpeech-2 (3) mix-noisy datasets.
The results shown in Table \ref{table:result} indicate that the DPRNN network achieves the best performance across all metrics, i.e., \ac{SI-SDRi} to 14.96 dB, \ac{SDRi} to 15.18 dB, PESQi to 0.930, and STOIi to 0.176 in separation tasks, maintaining great performance even on noisy datasets. Here, PIT association refers to the permutation invariant training strategy. This model serves as an upper bound, as ground-truth ID association is used.
However, its performance severely drops in speaker extraction tasks due to the absence of extraction cues. For the MMSpeech-2mix and -3mix datasets, the DPRNN network has achieved an accuracy of 50.73\% in the MMSpeech-2mix dataset and 17. 93\% in the MMSpeech-3mix dataset, respectively.

In the MMSpeech-2mix dataset, our proposed \ac{TPE} has achieved the best extraction performance, with \ac{SI-SDRi} of 12.16 dB, SDRi of 12.66 dB, PESQi of 0.830, STOIi of 0.150 and accuracy of 96.46\%. 
In the MMSpeech-3mix dataset, our proposed \ac{TPE} also achieved the best extraction performance, even better than the DPRNN network with the PIT association in all metrics except for STOIi. This shows that the text prompt is very instructive for speech extraction and also improves the quality of extraction.

Furthermore, we evaluated the performance of the SOTA sound extraction CLAPSep network~\cite{ma2024clapsep} finetuned on our dataset and the pretrained LASS network~\cite{liu22w_interspeech}. In finetune stage, we use the text prompt as positive query, and the interfering speech's text prompt as negative query. The results also demonstrate that these networks did not perform well on our \ac{TSE} task, highlighting the difficulty of using unaligned slide text cues.
LLM-TSE~\cite{hao2023typing} achieves \ac{SI-SDRi} of 10.97 dB and 10.30 dB on both MMSpeech-2(3) mix dataset, respectively. However, it still lags behind our TPE model by 1.19 dB and 1.64 dB on the same benchmarks.
Additionally, in terms of accuracy, LLM-TSE falls behind our TPE by 2.86\% and 5.00\% on both MMSpeech-2(3) mix dataset, respectively.

\vspace{-0.3cm}

\subsection{Ablation Study of \ac{TPE}}
\vspace{-0.1cm}
To validate the effectiveness of our key design choices, we conducted two ablation experiments. The results are shown in Table~\ref{tab:TSR_ablation}. First, we replaced our CLAP-based text encoder with a standard BERT model and used a simpler concatenation-based fusion. This change led to a clear performance degradation, with the \ac{SI-SDRi} dropping from 12.16 dB to 10.74 dB. More critically, when we removed only the \text{FiLM} fusion mechanism and replaced it with concatenation, the performance collapsed catastrophically: the \ac{SI-SDRi} plummeted to just 1.16 dB, and the accuracy dropped by over 24\% (from 96.46\% to 71.96\%). These results powerfully demonstrate that our chosen text encoder is effective, and that the \text{FiLM}-based fusion mechanism is absolutely essential for the success of our \ac{TPE} network.

\vspace{-0.3cm}
\subsection{\textbf{Robustness of Noise}}
\vspace{-0.1cm}

\begin{table}[]
    \centering
    \caption{The ablation study results.}
    \begin{adjustbox}{max width=0.45\textwidth}
    \begin{tabular}{c c c c c c}
        \toprule\toprule
         &Ablation &SI-SDRi (dB)$\uparrow$   &PESQi$\uparrow$    &STOIi$\uparrow$    &Acc (\%)$\uparrow$\\
         \midrule
         &TPE w/ BERT &10.74 &0.537 &0.110 &93.17 \\
         &TPE w/o FiLM &1.16 &-0.042 &-0.026 &71.96    \\
         &\textbf{TPE} & {\bf 12.16}       &{\bf 0.830}      & {\bf 0.150}     &{\bf 96.46} \\
         \bottomrule\bottomrule
    \end{tabular}
    \end{adjustbox}
\label{tab:TSR_ablation}
\end{table}

To evaluate the robustness of noise, we test our TPE model on several unseen noise datasets, including ESC-50~\cite{piczak2015dataset}, MS-SNSD~\cite{reddy2019scalable}, MUSAN~\cite{snyder2015musan}, Nonspeech~\cite{rashid2023nonspeech7k}, QUT-NOISE~\cite{dean2010qut}, and UrbanSound8k~\cite{salamon2014dataset}, with the results given in Table \ref{tab:SISDR_diff_noise}.
In particular, all noise segments are resampled to 16 kHz and then mixed 
at various \ac{SNR} levels for evaluation. 
As shown, TPE performs the best on the WHAM! noise dataset (which we used for training). 
When testing with unseen noise datasets, TPE shows the best performance on the MS-SNSD dataset and performs the worst on the QUT-NOISE dataset on average. 
The QUT-NOISE dataset contains numerous rapidly varying and non-stationary noises. As a result, \ac{TPE} exhibits worse performance on QUT-NOISE dataset compared to WHAM! dataset.
Despite this, the performance gap is not significantly large compared to the trained WHAM! dataset, demonstrating that our TPE exhibits strong robustness even under unseen noise data.

\vspace{-0.3cm}
\subsection{\textbf{Visualization of t-SNE}}
\vspace{-0.1cm}
To better explore the role of high-dimensional features of the text prompt in TSE, we perform t-SNE visualization~\cite{van2008visualizing} on the output of the final layer of the DPRNN block. 
We conduct the following experiments on the test and validation of MMSpeech-2mix dataset, including nine different speakers. Specifically, we randomly remove a portion of the text prompt with a specified masking ratio. The masking ratios used are 100\%, 80\%, 60\%, 40\%, 20\%, and 0\%, corresponding to subfigures (a) to (f) respectively, with color indices representing different speakers. To ensure stability, we run each experiment 10 times and average the results.

Additionally, the text prompt of the same speaker is expected to exhibit a certain level of similarity over a specific period of speech. It can be observed that when all text is removed, the SI-SDR is only -8.311 dB, and the corresponding t-SNE plot exhibits an almost random distribution. As the masking ratio decreases, the 80\% masked case still does not show clear clustering. However, when the masking ratio is reduced to 60\% or lower, distinct clustering patterns become apparent. This indicates that the text prompt is highly correlated with the target speaker which is significant for TSE.

\vspace{-0.3cm}
\section{Conclusion}
\vspace{-0.1cm}
In this work, we explore the use of unaligned text cues for \ac{TSE}.
Our proposal differs from previous works that have relied on various speaker cues (e.g, pre-registered speech, co-speech facial motions) which often face challenges in practical applications due to their 
complexity and intrusive nature. 
Specifically, we introduce a novel network architecture, namely \ac{TPE}, which excels in leveraging text prompts for \ac{TSE}.
The empirical results demonstrate the efficacy of our proposed methods, with the TPE network consistently outperforming existing methods across all metrics and evaluation conditions.
Moreover, we validate the noise robustness of \ac{TPE}  by evaluating on additional noise datasets. 
To gain deeper insights into how text prompts influence \ac{TPE}'s performance, we utilize t-SNE visualization to analyze their role in \ac{TSE}. The results reveal a strong correlation between the text prompt and the target speaker.
Furthermore,  we will make the dataset and source code publicly available to encourage further research and development in this field.

\vfill\pagebreak

\bibliographystyle{IEEEbib}
\bibliography{strings,refs}

\end{document}